\documentclass[aps,pre,twocolumn,superscriptaddress,floatfix]{revtex4}
\usepackage{graphicx}
\usepackage{bm}
\usepackage[dvips,unicode,colorlinks,linkcolor=blue,citecolor=blue,urlcolor=blue]{hyperref}

\begin{document}
\title{
Modeling of second sound in carbon nanostructures
}
\author{Alexander V. Savin}
\affiliation{Semenov Institute of Chemical Physics, Russian Academy of Sciences,
Moscow 119991, Russia}

\author{Yuri S. Kivshar}
\affiliation{Nonlinear Physics Center, Department of Fundamental and Theoretical Physics,
Research School of Physics, Australian National University, Canberra ACT 2601, Australia}

\date{\today}

\begin{abstract}
The study of thermal transport in low-dimensional materials has attracted a lot of attention
recently after discovery of high thermal conductivity of graphene. Here we study numerically
phonon transport in low-dimensional carbon structures being interested in the hydrodynamic 
regime revealed through the observation of second sound.  
We demonstrate that correct numerical modeling of such two-dimensional systems
requires semi-classical molecular dynamics simulations of temperature waves that take into
account quantum statistics of thermalized phonons. We reveal that second sound can be attributed
to the maximum group velocity of bending optical oscillations of carbon structures, and the 
hydrodynamic effects disappear for $T>200$K, being replaced by diffusive dynamics of thermal waves.
Our numerical results suggest that the velocity of second sound in such low-dimensional structures 
is about 6 km/s, and the hydrodynamic effects are manifested stronger in carbon nanotubes rather 
than in carbon nanoribbons.
\end{abstract}

\maketitle

\section{Introduction}
\label{s1}

Quasi-one-dimensional molecular systems such as carbon nanotubes and two-dimensional
atomic layers are known to possess many unusual physical properties. In particular,
thermal conductivity in such systems can be unexpectedly high exhibiting unusual phenomena
which are important for both fundamental physics and technological applications
of graphene and other two-dimensional materials \cite{Balandin2011,Nika2017,Gu2018,Zhang2020,Fu2020}.
One of such phenomena is associated with {\em second sound}, 
heat waves or hydrodynamic phonon transfer observed at temperatures above 100~K.

In three-dimensional materials, second sound was previously observed experimentally only 
at cryogenic temperatures~\cite{Ackerman1966,Jackson1972,Narayanamurti1972,Pohl1976,Hehlen1995},
as a reaction of a material to an applied temperature pulse.
Discovered later, exceptionally high thermal conductivities of low-dimensional materials (such as 
carbon nanotubes and nanoribbons, graphene, and boron nitride) occur due to a ballistic flow
of long-wave acoustic phonons supported by such systems~\cite{Lee2015,Nika2017,Yu2021,Zhang2021,Liu2021,Sachat2021}.
Second sound, or hydrodynamic phonon transfer, is observed between ballistic and diffusion regimes 
of heat transfer.
In two-dimensional (2D) materials, second sound can be observed at temperatures above 100K 
\cite{Huberman2019,Ding2018,Ding2022}, and for rapidly changing temperature high-frequency second sound
can also be observed in three-dimensional (3D) materials at higher temperatures~\cite{Beardo2021}.

To describe the hydrodynamic regime of phonon transfer in low-dimensional materials, several theoretical
models were proposed~\cite{Cepellotti2015,Lee2015,Ding2018,Luo2019,Shang2020,Yu2021,Chiloyan2021,Scuracchio2019}.
In particular, a 3D model of a periodically modulated graphene structure that behaves like a crystal
for temperature waves was proposed in Ref.~\cite{Gandolfi2020}. 
Hydrodynamic features of the phonon transfer in a single-wall carbon nanotube with chirality 
indices (20,20) were discussed in Ref.~\cite{Lee2017}, where the authors suggested a formula for 
the contribution of phonon drift motion into the total heat flow, and the second sound velocity
in the nanotube was estimated as $v_s=4$km/s. 
The second sound velocity in the graphene was estimated  in Refs.~\cite{Lee2015,Scuracchio2019} as
$v_s=3.2$km/s (for temperature 100K).
\begin{figure}[t]
\includegraphics[angle=0, width=1.0\linewidth]{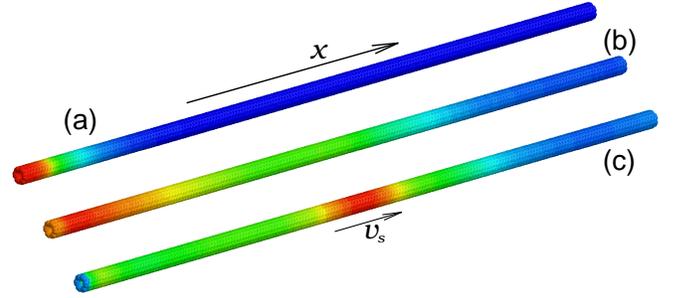}
\caption{
Motion of a heat pulse along a carbon nanotube with chirality index (6,6).
Case (a) is the initial condition at $t=0$. Cases (b,c) show schematically the temperature distribution
along the nanotube for two different scenarios of heat transfer: (b) spreading, for high temperature
$T=300$K, (c) propagation of a heat wave, for low temperature $T=50$, where $v_s$ is the speed of 
second sound.
}
\label{fig00}
\end{figure}

Direct numerical simulation of second sound in solids is a difficult task because
one needs to study polyatomic molecular systems also taking into account quantum statistics of phonons.
Propagation of temperature pulses was simulated for single-wall
\cite{Osman2005,Shiomi2006,Chen2011,Mashreghi2011} and double-wall carbon nanotubes 
(CNTs)~\cite{Gong2013} and carbon nanoribbons (CNRs)~\cite{Yao2014},
by employing the classical method of molecular dynamics. This approach predicts thermalization 
of all phonons regardless of their frequency and temperature. 
The thermal pulse was generated by connecting a short edge section of a nanotube (or a nanoribbon) 
with a thermostat at temperature $T_h=800K$, or $1000K$ for a short time ($\sim$1~ps).
Either a zero value \cite{Osman2005,Chen2011,Mashreghi2011,Gong2013} or $T=50K$
\cite{Shiomi2006,Yao2014} was used as a background temperature. The motion of a thermal pulse
along the nanotube was analyzed through the study of spatiotemporal temperature profiles.
It  was shown that the initial thermal pulse causes the formation of several wave packets,
the leading one moving at the speed of long-wave acoustic phonons.

In this paper, we study the propagation of thermal pulses along carbon nanotubes and carbon nanoribbons
(see Fig.~\ref{fig00}) by using semi-classical molecular dynamic simulations~\cite{Savin2012}.
This approach allows us to model thermalized phonons also taking into account their quantum statistics,
i.e. taking into account the full thermalization of low-frequency phonons with frequencies $\omega<k_BT/\hbar$
and partial thermalization of phonons with frequencies $\omega>k_BT/\hbar$, where $k_B$ is the Boltzmann
constant, $\hbar$ is the Planck constant. As we demonstrate below,  our approach allows to simulate
second sound in graphene at $T\approx 100$K only taking into account quantum statistics of phonons.
In that way, we can study not only the propagation of a short thermal pulse but also
the dynamics of periodic sinusoidal temperature profiles, as relaxation of the periodic 
temperature lattices.

The paper is organized as follows. Section~\ref{s2} describes our full-atomic model of carbon nanoribbons
and carbon nanotubes, which is further employed to simulate numerically the heat transport.
In Sec.~\ref{s3}, we construct the dispersion curves of nanoribbons and nanotubes,
and analyze their characteristics and the propagation velocities of low-frequency phonons. 
Section~\ref{s4} describes our method of semi-classical molecular dynamics simulations. 
Then in Sec.~\ref{s5}, we simulate numerically different regime of the propagation of a thermal pulse,
and in Sec.~\ref{s6} we study relaxation of temperature periodic lattices.
More specifically, we demonstrate the existence of second sound for temperatures $50\div 150$K.
In Sec.~\ref{s7}, for a direct comparison, we study the dynamics of thermal pulses and relaxation
of periodic temperature lattices by employing the classical molecular dynamics.
Section \ref{s8} concludes our paper. Thus, we reveal that the quantum statistics of phonons
should be taken into account for the study of second sound in low-dimensional systems.

\section{Model}
\label{s2}
\begin{figure}[t]
\includegraphics[angle=0, width=1\linewidth]{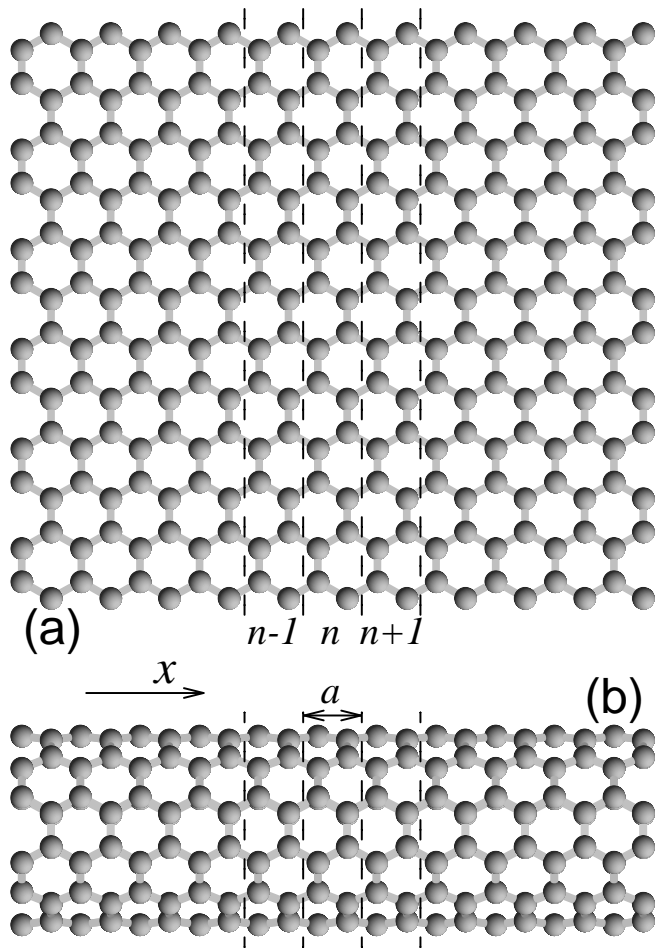}
\caption{
Atomic model of (a) carbon zigzag nanoribbon and (b) nanotube with chirality index (6,6).
Nanoribbon and nanotube are placed along the $x$ axis, $a$ is longitudinal step, $n$ is number of transverse
unit cells (dashed lines separate unit cells), and the number atoms in the unit cell $K=24$.
}
\label{fig01}
\end{figure}

We consider a planar carbon nanoribbon (CNR) and a nanotube (CNT) with a zigzag
structure  consisting of $N\times K$ atoms -- see Fig. \ref{fig01} ($N$ is the number
of transverse unit cells, $K$ -- the number of atoms in the unit cell).
The nanoribbon is assumed to be flat in the ground state. Initially, we assume that the
nanoribbon lies in the $xy$ plane, and its symmetry center is directed along the $x$ axis.
Then its length can be calculated as $L_x=(N-0.5)a$, width
$L_y=3Kr_0/4-r_0$, where the longitudinal step of the nanoribbon is $a=r_0\sqrt{3}$,
$r_0 = 1.418$~\AA~-- C--C valence bond length.

In realistic cases, the edges of the nanoribbon are always chemically modified.
For simplicity, we assume that the hydrogen atoms are attached to each edge carbon atom
forming the edge line of CH groups. In our numerical simulations, we take this into account
by a change of the mass of the edge atoms. We assume that the edge carbon atoms have
the mass $M_1=13m_p$, while all other internal carbon atoms have the mass $M_0=12m_p$,
where $m_p=1.6601\times 10^{-27}$~kg is the proton mass.

\begin{figure}[t]
\includegraphics[angle=0, width=1\linewidth]{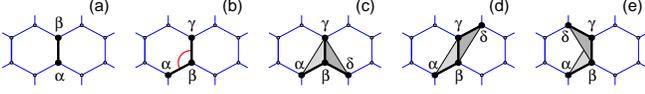}
\caption{
Configurations containing up to $i$th type of nearest-neighbor interactions for
(a) $i=1$, (b) $i=2$, (c) $i=3$, (d) $i=4$, and (e) $i=5$.
}
\label{fig02}
\end{figure}

Hamiltonian of the nanoribbon and nanotube can be presented in the form,
\begin{equation}
H=\sum_{n=1}^N\sum_{k=1}^K\big[\frac12M_{n,k}(\dot{\bf u}_{n,k},\dot{\bf u}_{n,k})+P_{n,k}\big],
\label{f1}
\end{equation}
where each carbon atom has a two-component index $\alpha=(n,k)$, $n$ is the number of transversal
elementary cell of zigzag nanoribbon (nanotube), $k$ is the number of atoms in the cell.
Here $M_\alpha$ is the mass of the carbon atom with the index $\alpha$ (for internal atoms
of nanoribbon and for all atoms of nanotube, $M_\alpha=M_0$, for the edge atoms of nanoribbon,
$M_\alpha=M_1$), ${\bf u}_\alpha=(x_\alpha(t),y_\alpha(t),z_\alpha(t))$ is the three-dimensional
vector that describes the position of an atom with the index $\alpha$ at the time moment $t$.
The term $P_\alpha$ describes the interaction of the carbon atom with the index $\alpha$
with the neighboring atoms. The potential depends on variations
in bond length, in bond angles, and in dihedral angles between the planes formed by three
neighboring carbon atoms. It can be written in the form
\begin{equation}
P=\sum_{\Omega_1}U_1+\sum_{\Omega_2}U_2+\sum_{\Omega_3}U_3+\sum_{\Omega_4}U_4+\sum_{\Omega_5}U_5,
\label{f2}
\end{equation}
where $\Omega_i$, with $i=1$, 2, 3, 4, 5, are the sets of configurations including all interactions
of neighbors. These sets only need to contain configurations of the atoms shown in Fig.~\ref{fig02},
including their rotated and mirrored versions.

Potential $U_1({\bf u}_\alpha,{\bf u}_\beta)$ describes the deformation energy due to a direct interaction
between pairs of atoms with the indexes $\alpha$ and $\beta$, as shown in Fig.~\ref{fig02}(a).
The potential $U_2({\bf u}_\alpha,{\bf u}_\beta,{\bf u}_\gamma)$
describes the deformation energy of the angle between the valence bonds
${\bf u}_\alpha{\bf u}_\beta$, and ${\bf u}_\beta{\bf u}_\gamma$, see Fig.~\ref{fig02}(b).
Potentials $U_i({\bf u}_\alpha,{\bf u}_\beta,{\bf u}_\gamma,{\bf u}_\delta)$,
$i=3$, 4, and 5, describe the deformation energy associated with a change in the angle between the
planes ${\bf u}_\alpha{\bf u}_\beta{\bf u}_\gamma$ and ${\bf u}_\beta{\bf u}_\gamma{\bf u}_\delta$,
as shown in Figs.~\ref{fig02}(c)--(e).

We use the potentials employed in the modeling of the dynamics of large polymer macromolecules
\cite{Noid1991,Sumpter94} for the valence bond coupling,
\begin{equation}
U_1({\bf u}_1,{\bf u}_2)\!=\!\epsilon_1
\{\exp[-\alpha_0(\rho-\rho_0)]-1\}^2,~\rho\!=\!|{\bf u}_2-{\bf u}_1|,
\label{f3}
\end{equation}
where $\epsilon_1=4.9632$~eV is the energy of the valence bond and $\rho_0=1.418$~\AA~
is the equilibrium length of the bond;
the potential of the valence angle
\begin{eqnarray}
U_2({\bf u}_1,{\bf u}_2,{\bf u}_3)=\epsilon_2(\cos\varphi-\cos\varphi_0)^2,~~
\label{f4}\\
\cos\varphi=({\bf u}_3-{\bf u}_2,{\bf u}_1-{\bf u}_2)/
(|{\bf u}_3-{\bf u}_2|\cdot |{\bf u}_2-{\bf u}_1|),~~
\nonumber
\end{eqnarray}
so that the equilibrium value of the angle is defined as $\cos\varphi_0=\cos(2\pi/3)=-1/2$;
the potential of the torsion angle
\begin{eqnarray}
\label{f5}
U_i({\bf u}_1,{\bf u}_2,{\bf u}_3,{\bf u}_4)=\epsilon_i(1+z_i\cos\phi),\\
\cos\phi=({\bf v}_1,{\bf v}_2)/(|{\bf v}_1|\cdot |{\bf v}_2|),\nonumber \\
{\bf v}_1=({\bf u}_2-{\bf u}_1)\times ({\bf u}_3-{\bf u}_2), \nonumber \\
{\bf v}_2=({\bf u}_3-{\bf u}_2)\times ({\bf u}_3-{\bf u}_4), \nonumber
\end{eqnarray}
where the sign $z_i=1$ for the indices $i=3,4$ (equilibrium value of the torsional angle $\phi_0=\pi$)
and $z_i=-1$ for the index $i=5$ ($\phi_0=0$).

The specific values of the parameters are $\alpha_0=1.7889$~\AA$^{-1}$,
$\epsilon_2=1.3143$ eV, and $\epsilon_3=0.499$ eV, they are found from the frequency
spectrum of small-amplitude oscillations of a sheet of graphite~\cite{Savin08}.
According to previous study~\cite{Gunlycke08}, the energy $\epsilon_4$ is close to
the energy $\epsilon_3$, whereas  $\epsilon_5\ll \epsilon_4$
($|\epsilon_5/\epsilon_4|<1/20$). Therefore, in what follows we use the values
$\epsilon_4=\epsilon_3=0.499$ eV and assume $\epsilon_5=0$, the latter means that we
omit the last term in the sum (\ref{f2}).
More detailed discussion and motivation of our choice of the interaction potentials
(\ref{f3}), (\ref{f4}), (\ref{f5}) can be found in earlier publication~\cite{Savin10}.

\section{Dispersion curves}
\label{s3}
Let us consider carbon nanoribbon (nanotube) in the equilibrium state
$\{ {\bf u}_{n,k}^0\}_{n=-\infty,k=1}^{+\infty,K}$ which is characterised by longitudinal shift $a$
and by the positions of $K$ atoms in the elementary cell: ${\bf u}_{n,k}^0={\bf u}_{0,k}^0+an{\bf e_x}$,
where vector $e_x=(1,0,0)$ -- see Fig.~\ref{fig01}.

Then, we introduce $3K$-dimensional vector, ${\bf v}_n=\{ {\bf u}_{n,k}-{\bf u}_{n,k}^0\}_{k=1}^K$,
that describes a shift of the atoms  of the $n$th cell from its equilibrium positions.
The nanoribbon (nanotube) Hamiltonian can be written in the following form:
\begin{equation}
H=\sum_n\{\frac12({\bf M}\dot{\bf v}_n,\dot{\bf v}_n)+P({\bf v}_{n-1},{\bf v}_{n},{\bf v}_{n+1})\},
\label{f6}
\end{equation}
where ${\bf M}$ is the diagonal matrix of masses of all atoms of the elementary cell.

Hamiltonian (\ref{f6}) generates the following set of the equation of motion:
\begin{eqnarray}
-{\bf M}\ddot{\bf v}_n&=&P_1({\bf v}_{n},{\bf v}_{n+1},{\bf v}_{n+2})
+P_2({\bf v}_{n-1},{\bf v}_{n},{\bf v}_{n+1}) \nonumber\\
&&+P_3({\bf v}_{n-2},{\bf v}_{n-1},{\bf v}_{n}), \label{f7}
\end{eqnarray}
where function
$$
P_i({\bf v_1},{\bf v}_2,{\bf v}_3)=\frac{\partial}{\partial {\bf v}_i}P({\bf v_1},{\bf v}_2,{\bf v}_3),~i=1,2,3.
$$
In the linear approximation, this system takes the form
\begin{equation}
-{\bf M}\ddot{\bf v}_n={\bf B}_1{\bf v}_{n}+{\bf B}_2{\bf v}_{n+1}+{\bf B}_2^*{\bf v}_{n-1}
+{\bf B}_3{\bf v}_{n+2}+{\bf B}_3^*{\bf v}_{n-2},
\label{f7}
\end{equation}
where the matrix elements are defined as
$$
{\bf B}_1=P_{1,1}+P_{2,2}+P_{3,3},~{\bf B}_2=P_{1,2}+P_{2,3},~{\bf B}_3=P_{1,3}
$$
and the matrix of the partial derivatives takes the form
$$
P_{i,j}=\frac{\partial^2 P}{\partial{\bf v}_i,\partial{\bf v}_j}({\bf 0},{\bf 0},{\bf 0}),~~i,j,=1,2,3.
$$
Solution of the system linear equations (\ref{f7}) can be written in the standard form of the wave
\begin{equation}
{\bf v}_n=A{\bf w}\exp(iqn-i\omega t), \label{f8}
\end{equation}
where $A$ -- amplitude, ${\bf w}$ -- eigenvector, $\omega$ is phonon frequency with the dimensionless
wave number $q\in[0,\pi]$.
Substituting Eq. (\ref{f8}) into Eq. (\ref{f7}), we obtain the eigenvalue problem
\begin{equation}
\omega^2{\bf Mw}={\bf C}(q){\bf w}, \label{f9}
\end{equation}
where Hermitian matrix
$$
{\bf C}(q)={\bf B}_1+{\bf B}_2e^{iq}+{\bf B}_2^*e^{-iq}
+{\bf B}_3 e^{2iq}+{\bf B}_3 e^{-2iq}.
$$
Using the substitution ${\bf w}={\bf M}^{-1/2}{\bf e}$, problem (\ref{f9}) can be rewritten in the form
\begin{equation}
\omega^2{\bf e}={\bf M}^{-1/2}{\bf C}(q){\bf M}^{-1/2}{\bf e}, \label{f10}
\end{equation}
where ${\bf e}$ is the normalized eigenvector, $({\bf e},{\bf e})=1$.
\begin{figure}[t]
\includegraphics[angle=0, width=1\linewidth]{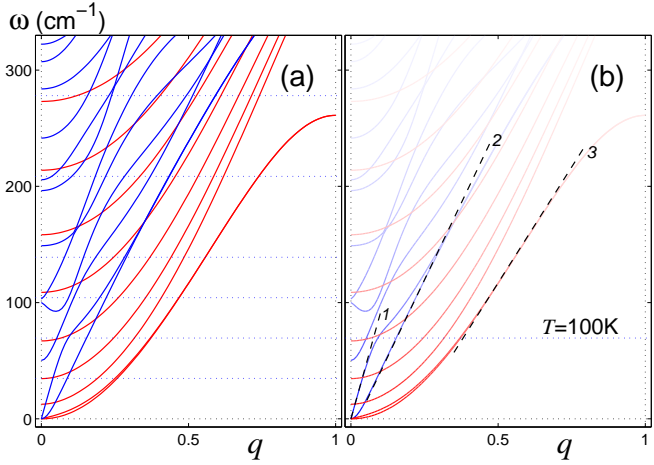}
\caption{
(a) Structure of dispersion curves
$\{ \omega_i(q) \}_{i=1}^{3K}$ of the zigzag nanoribbon
(the number atoms in transverse unit cell $K=24$).
The blue curves correspond  to the in-plane vibrations, whereas the red curves
correspond to out-to-plane vibrations.
Horizontal point lines show the values of the thermalization frequency of nonzero fluctuations
$\Omega(T)$ at temperatures $T=50$, 100, 150, 200, 300 and 400K.
(b) The degree of thermalization of nonzero oscillations
$p(\omega,T)$ at temperature $T=100$K
is shown through the intensity of the line color (full saturation for $p\approx 1$,  low
saturation for $p\approx 0$).
Dashed straight lines 1, 2, 3 set tangents to the first four dispersion curves, which
correspond to  wave velocities $v_o=5.88$, $v_t=7.79$, $v_l=13.36$~km/s.
}
\label{fig03}
\end{figure}

Therefore, in order to find the dispersion relations characterizing the modes of the nanoribbon (nanotube)
for each fixed value of the dimensionless wave number $0\le q\le\pi$ we need to find numerically
the eigenvalues of the Hermitian matrix [Eq. (\ref{f10})] of the order $3K\times 3K$.
As a result, we obtain 3K branches of the dispersion curve $\{ \omega_j(q)\}_{j=1}^{3K}$.

The plain structure of the nanoribbon allows us to divide its vibrations into two classes: into
in-plane vibrations, when the atoms always stay in the plane of the nanoribbon
and into out-of-plane vibrations when the atoms are shifted orthogonal to the plane.
Two third of the branches correspond to the atom vibrations
in the $xy$ plane of the nanoribbon (in-plane vibrations), whereas only one third corresponds to the
vibrations orthogonal to the plane (out-of-plane vibrations), when the atoms are shifted along the axes
$z$. The maximal frequency of in plane vibrations is $\omega_m=1598$~cm$^{-1}$, the maximum frequency
of out-of-plane vibrations is $\omega=898$~cm$^{-1}$.
This values goes in accord with the experimental data for planar graphite
\cite{AlJishi1982,Aizawa1990,Maultzsch2004}.
\begin{figure}[t]
\includegraphics[angle=0, width=1\linewidth]{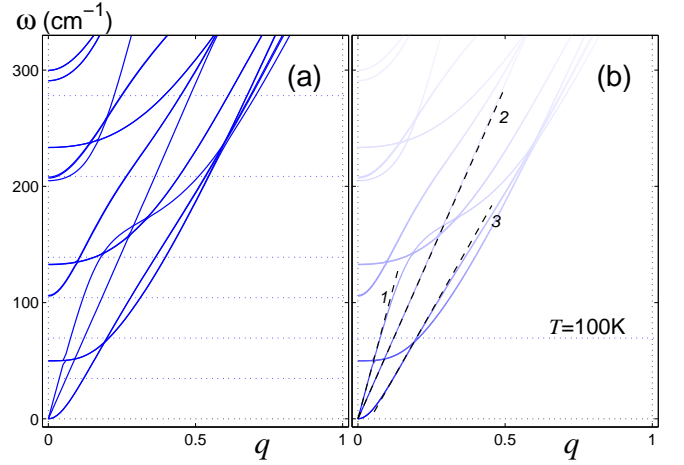}
\caption{
(a) Structure of dispersion curves
$\{\omega_i(q)\}_{i=1}^{3K}$ of nanotube
with chirality index (6,6).
Horizontal point lines show the values of the thermalization frequency of nonzero fluctuations
$\Omega(T)$ at temperatures $T=50$, 100, 150, 200, 300 and 400K.
(b) The degree of thermalization of nonzero oscillations
$p(\omega,T)$ at temperature $T=100$K
is shown through the intensity of the line color (full saturation for $p\approx 1$,  low
saturation for $p\approx 0$).
Dashed straight lines 1, 2, 3 set tangents to the first four dispersion curves, which
correspond to  wave velocities $v_o=6.54$, $v_t=8.41$, $v_l=13.84$~km/s.
}
\label{fig04}
\end{figure}

The form of nanoribbon dispersions curves in the low frequency region is shown in Fig.~\ref{fig03}.
Four branches of the curves start from the zero point ($q=0$, $\omega=0$).
Two first branches $\omega_i(q)$, $i=1,2$ correspond to the orthogonal (out-of-plane)
bending vibrations of the nanoribbon; third branch $\omega_3(q)$ describes the bending
planar (in-plane) vibrations in the plane. These branches approach smoothly the axis $q$:
$\omega_i(q)/q\rightarrow 0$, when $q\rightarrow 0$, so that the corresponding long-wave phonon
possess zero dispersion.
However, we can determine for them the maximum values of group velocities
$$
s_i=a\max_{0\le q\le\pi} d\omega_i(q)/dq,~~i=1,2,3.
$$
Forth branch $\omega_4(q)$ correspond to in-plane longitudinal vibrations.
The corresponding  long-wave mode posses nonzero dispersion so that we can define limiting value
$$
v_l=a\lim_{q\rightarrow 0} \omega_4(q)/q=13.36~\mbox{km/s},
$$
which define the sound speed of longitudinal
acoustic waves (phonons) of the nanoribbon.
The speeds of optical (bending) phonons will determine the values
$v_o=s_1,s_2=5.88$~km/s for out-of-plane and $v_t=s_3=7.79$~km/s for in-plane vibrations
-- see Fig.~\ref{fig03} (b).

The maximal frequency of nanotube vibrations $\omega_m=1630$~cm$^{-1}$. The form of dispersion curve
in the low frequency region is shown in Fig.~\ref{fig04}. For nanotube four branches of the curve
also start from zero point. Two first branches $\omega_i(q)$, $i=1,2$ correspond to the bending vibrations.
These branches approach smoothly the axis $q$: $\omega_i(q)/q\rightarrow 0$, when $q\rightarrow 0$
(corresponding long-wave binding phonon posses zero dispersion). For them we can determine the
maximum values of group  velocities
$$
v_o=a\max_{q\in[0,\pi],i=1,2} d\omega_i(q)/dq=6.54~\mbox{km/s}.
$$
Third and forth branches correspond to torsional and longitudinal acoustic phonon of the nanotube.
The corresponding long-wave modes posses nonzero dispersion so that we can define limiting values
$$
s_{i}=a\lim_{q\rightarrow 0} \omega_{i}(q)/q,~~i=3,4,
$$
which define the sound speed of acoustic
torsional $v_t=s_3=8.41$~km/s and longitudinal waves $v_l=s_4=13.84$~km/s -- see Fig.~\ref{fig04} (b).

Let us note that when taking into account the quantum statistics of thermal phonons,
only nonzero phonons with the average energy of $k_BTp(\omega,T)$ can participate
in heat transfer, where the degree of thermalization
\begin{equation}
p(\omega,T)=\frac{\hbar\omega/k_BT}{\exp(\hbar\omega/k_BT)-1}, \label{f12}
\end{equation}
depends on temperature $T$ and phonon frequency $\omega$ (function $0<p(\omega,T)<1$,
$k_B$ and $\hbar$ are the Boltzmann and Plank constants) \cite{Landau}.
Here it is taken into account that zero-point oscillations are not involved
in the heat phonon transport.

As the temperature increases, $T\nearrow\infty$, density $p\nearrow 1$.
At high temperatures, all phonons become equally thermalized, each has the average energy equal
to $k_BT$ (classical approximation).
When the temperature decreases, high-frequency nonzero phonons freeze out ($p(\omega,T)\searrow 0$
when $T\searrow 0$), only phonons with frequencies $\omega<\Omega(T)$
where thermalization frequency
$$
\Omega(T)=k_BT/\hbar,
$$
remain thermalized [function $p(\Omega(T),T)=1/(e-1)=0.582$].
Therefore, at low temperatures, only low-frequency phonons will participate in heat transfer.
Thus, at $T<400$K, phonons with frequencies $\omega<300$~cm$^{-1}$ will participate in heat transfer,
and at $T<50$K it will be only low-frequency long-wave phonons (see Figs.~\ref{fig03} and \ref{fig04}).
The degree of participation of a phonon with a frequency of $\omega$ in heat transfer
is characterized by the function (\ref{f12}).

\section{Interaction with a thermostat}
\label{s4}
In the classical approach interaction of nanoribbons (nanotubes) with a thermostat is described
by the Langevin system of equations
\begin{equation}
{\bf M}\ddot{\bf x}_n=-\frac{\partial H}{\partial{\bf x}_n}
-\gamma{\bf M}\dot{\bf x}_n+\Xi_n,~~1\le n\le N,
\label{f13}
\end{equation}
where $\bf M$ is $3K\times 3K$-dimensional diagonal mass matrix,
$3K$-dimensional vector ${\bf x}_n=\{ {\bf u}_{n,k}\}_{k=1}^K$ gives the coordinates of
carbon atoms from $n$th transverse unit cell,
damping coefficient $\gamma=1/t_0$ ($t_0$ -- relaxation time) and $\Xi_n=\{\xi_{n,k,i}\}_{k=1,i=1}^{K,~3}$
is $3K$-dimensional vector of normally distributed random forces (white noise) normalized by conditions
\begin{equation}
\langle\xi_{\alpha,i}(t_1)\xi_{\beta,j}(t_2)\rangle
= 2M_{\alpha}\gamma k_BT\delta_{\alpha\beta}\delta_{ij}\delta(t_1-t_2).
\label{f14}
\end{equation}

In the semiquantum approach the random forces do not represent in general white noise.
The power spectral density of the random forces in that description should be given by the
quantum fluctuation-dissipation theorem \cite{Landau,Callen1951}:
\begin{equation}
\langle\xi_{\alpha,i}\xi_{\beta,j}\rangle_\omega
= 2M_{\alpha}\gamma k_BT\delta_{\alpha\beta}\delta_{ij}p(\omega,T).
\label{f15}
\end{equation}

To model the nanoribbon (nanotube) stochastic dynamics in the semiquantum approach \cite{Savin2012},
we will use the Langevin equations of motions (\ref{f13}) with random forces
$\Xi_n=\{\zeta_{n,k,i}\}_{k=1,i=1}^{K,~3}$ with the power spectral density,
given by $p(\omega,T_n)$, where $T_n$ is temperature of the $n$th transverse unit cell.
This dimension color noise $\zeta_{n\alpha}$ is conveniently  derived from the dimensionless noise
$S_{n\alpha}(\tau)$: $\zeta_{n\alpha}(t)=k_BT_n\sqrt{2M_{n\alpha}\gamma/\hbar}S_{n\alpha}(\tau)$,
where dimensionless time $\tau=k_BT_nt/\hbar$, spectral density
\begin{equation}
p_1(\bar\omega)=\bar\omega/(\exp(\bar\omega)-1),
\label{f16}
\end{equation}
dimensionless frequency $\bar\omega=\hbar\omega/k_BT_n$.

The random function $S_{n\alpha}(\tau)$, which will generate the power spectral density $p_1(\bar\omega)$,
can be approximated by a sum of two random functions with narrow frequency spectra:
\begin{equation}
S_{n\alpha}(\tau)=c_1\varsigma_{1,n\alpha}(\tau)+c_2\varsigma_{2,n\alpha}(\tau).
\label{f17}
\end{equation}
In this sum the dimensionless random functions $\varsigma_{i,n\alpha}(\tau)$, $i=1,2$,
satisfy the equations of motion as
\begin{equation}
\varsigma_{i,n\alpha}''(\tau)=\eta_{i,n\alpha}(\tau)-\bar\Omega_i^2\varsigma_{i,n\alpha}(\tau)
-\bar\Gamma_i\varsigma_{i,n\alpha}'(\tau),
\label{f18}
\end{equation}
where $\eta_{i,n\alpha}(\tau)$ are $\delta$-correlated white-noise functions:
$$
\langle\eta_{i,n\alpha}(\tau)\eta_{j,k\beta}(0)\rangle=2\bar\Gamma_i\delta_{ij}\delta_{nk}\delta_{\alpha\beta}\delta(\tau).
$$
The power spectral density of the sum of two random functions (\ref{f17})
$\langle S_{n\alpha}S_{k\beta}\rangle_{\bar\omega}=\delta_{nk}\delta_{\alpha\beta}p_2(\bar\omega)$,
where function
\begin{equation}
p_2(\bar\omega)=\sum_{i=1}^2\frac{2c_i^2\bar\Gamma_i}{(\bar\Omega_i^2-\bar\omega^2)^2+\bar\omega^2\bar\Gamma_i^2}.
\label{f19}
\end{equation}
The function $p_2(\bar\omega)$ approximate with high accuracy the function $p_1(\bar\omega)$
for the values of dimensionless parameters
$c_i$, $\bar\Omega_i$, $\bar\Gamma_i$, $i=1,2$ represented in Table \ref{tab1}.
\begin{table}[tb]
\centering\noindent
\caption{Value of the coefficients $c_i$, $\bar\Omega_i$, $\bar\Gamma_i$.
}
\label{tab1}
\begin{tabular}{cccccc}
\hline
\hline
$c_1$  & $c_2$  & $\bar\Omega_1$ & $\bar\Omega_2$ & $\bar\Gamma_1$ & $\bar\Gamma_2$\\
1.8315 & 0.3429 & 2.7189         & 1.2223         & 5.0142         & 3.2974\\
\hline
\hline
\end{tabular}
\end{table}

Thus, in order to obtain the thermalized state of the nanoribbon (nanotube), it is necessary to solve
numerically the system of equation of motion with color noise:
\begin{equation}
{\bf M}\ddot{\bf x}_n=-\frac{\partial H}{\partial{\bf x}_n}
-\gamma{\bf M}\dot{\bf x}_n+c_1\Theta_{1,n}+c_1\Theta_{2,n},~~1\le n\le N,
\label{f20}
\end{equation}
where $3K$-dimensional vector of random forces
$\Theta_{i,n}=\{\zeta_{i,n,k,j}\}_{k=1,j=1}^{K,~3}$, $i=1,2$,
is a solution of system of linear equations
\begin{equation}
\ddot{\zeta}_{i,n,\alpha}\!=\! \eta_{i,n,\alpha}-\left(\bar{\Omega}_i\frac{k_BT_n}{\hbar}\right)^2\!\!\zeta_{i,n,\alpha}
-\bar{\Gamma}_i\frac{k_BT_n}{\hbar}\dot{\zeta}_{i,n,\alpha},~i=1,2,
\label{f21}
\end{equation}
where $\eta_{i,n,\alpha}$ are normally distributed random forces (white noise)
normalized by conditions
\begin{equation}
\langle\eta_{i,n,\alpha}(t)\eta_{j,k,\beta}(0)\rangle\!\!
=\!4\bar{\Gamma}_i\left(\frac{k_BT_n}{\hbar}\right)^5\!\!\hbar\gamma M_\alpha\delta_{ij}\delta_{nk}\delta(t).
\label{f22}
\end{equation}

The system of equations (\ref{f20}), (\ref{f21}) was integrated numerically with the initial conditions
\begin{eqnarray}
{\bf x}_n(0)={\bf x}_n^0,~\dot{\bf x}_n(0)={\bf 0},~\Theta_{i,n}={\bf 0},
~\dot{\Theta}_{i,n}={\bf 0}
\label{f23}\\
n=1,2,...,N,~~i=1,2,\nonumber
\end{eqnarray}
where ${\bf x}_n^0$ is the coordinate of carbon atoms in ground stationary state of nanoribbon (nanotube).
The value of the relaxation time $t_0$ characterizes the intensity of the exchange
of the molecular system with the thermostat.
To achieve the equilibrium of the system with the thermostat,
it is enough to integrated the system of equations of motion during the time $t=10t_0$.

In the simulation, the value $t_0=0.4$~ps was used.
The system equations (\ref{f20}), (\ref{f21}) was integrated  numerically during the time $t_1=5$~ps.
Further the interaction with the thermostat was turned off, i.e. the system of equations of motion
without friction and random forces
\begin{equation}
{\bf M}\ddot{\bf x}_n=-\frac{\partial H}{\partial {\bf x}_n},~n=1,2,...,N,
\label{f24}
\end{equation}
was numerically integrated.

Without taking into account zero-point oscillations, the normalized kinetic energy of thermal phonons
\begin{equation}
e(T)=\frac{1}{3NK-6}\sum_{n=7}^{3NK}\frac{\hbar\Omega_n}{\exp(\hbar\Omega_n/k_BT)-1},
\label{f25}
\end{equation}
where $\{\Omega_n\}_{n=7}^{3NK}$ are the nonzero natural oscillations frequencies of the nanoribbon
(nanotube). The average value of kinetic energy can also be obtained from the integration of a
thermalized system of equations of motion (\ref{f24})
\begin{eqnarray}
e_{kin}=\frac{1}{3NK}\langle\sum_{n=1}^N({\bf M}\dot{\bf x}_n,\dot{\bf x}_n)\rangle
\nonumber\\
=\frac{1}{3NKt}\int_0^t\sum_{n=1}^N\left({\bf M}\dot{\bf x}_n(s),\dot{\bf x}_n(s)\right)ds.
\label{f26}
\end{eqnarray}
For a uniformly thermalized molecular system, these energies must coincide:
\begin{equation}
e(T)=e_{kin}.
\label{f27}
\end{equation}
Since function $e(T)$ monotonically increases with $T$, Eq. (\ref{f27}) has a unique solution
for the temperature. For inhomogeneous thermalization, the solution of the equation
\begin{equation}
e(T_n)=\frac{1}{3K}\langle({\bf M}\dot{\bf x}_n,\dot{\bf x}_n)\rangle
\label{f28}
\end{equation}
allows us to find the distribution of temperature along the nanoribbon (nanotube)  $\{T_n\}_{n=1}^N$.

In classical approach, i.e. by using a system of Langevin equations (\ref{f13}) with white noise
(\ref{f14}), the temperature distribution can be obtained from the formula
\begin{equation}
T_n=\frac{1}{3Kk_B}\langle({\bf M}\dot{\bf x}_n,\dot{\bf x}_n)\rangle
\label{f29}
\end{equation}
\begin{figure}[t]
\includegraphics[angle=0, width=1.0\linewidth]{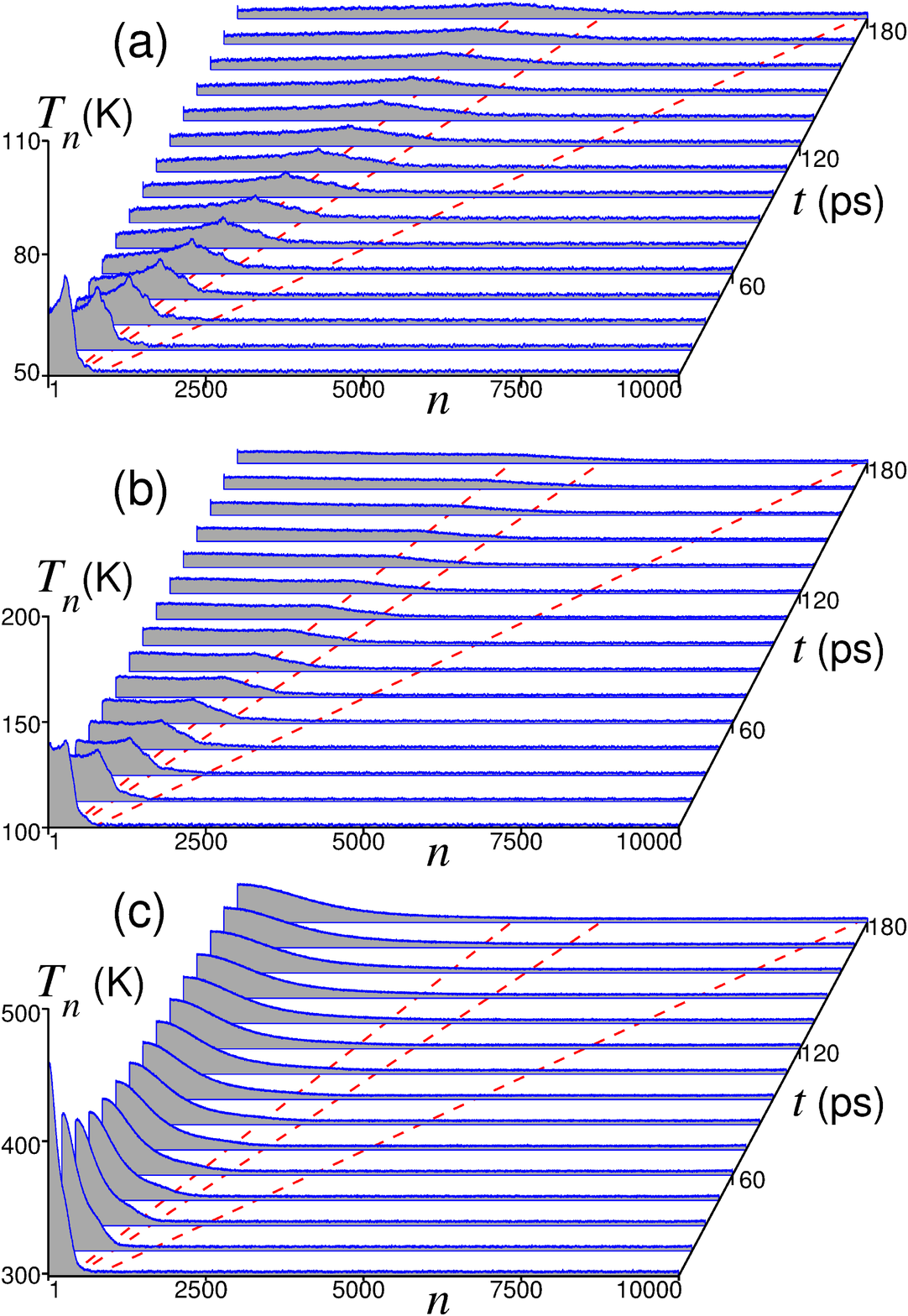}
\caption{
Dynamics of the temperature impulse along nanoribbon (width $L_y=2.41$~nm)
at its initial temperature $T_h=3T$, length $N_t=40$ and background temperature
(a) $T=50$, (b) 100 and (c) 300 K.
The dependence of the temperature distribution along the CNR $T_n$ on the time $t$ is shown.
The dashed (red) lines shows trajectories for moving with constant speed $v=v_o$, $v_t$ and $v_l$.
}
\label{fig05}
\end{figure}
\begin{figure}[t]
\includegraphics[angle=0, width=1.0\linewidth]{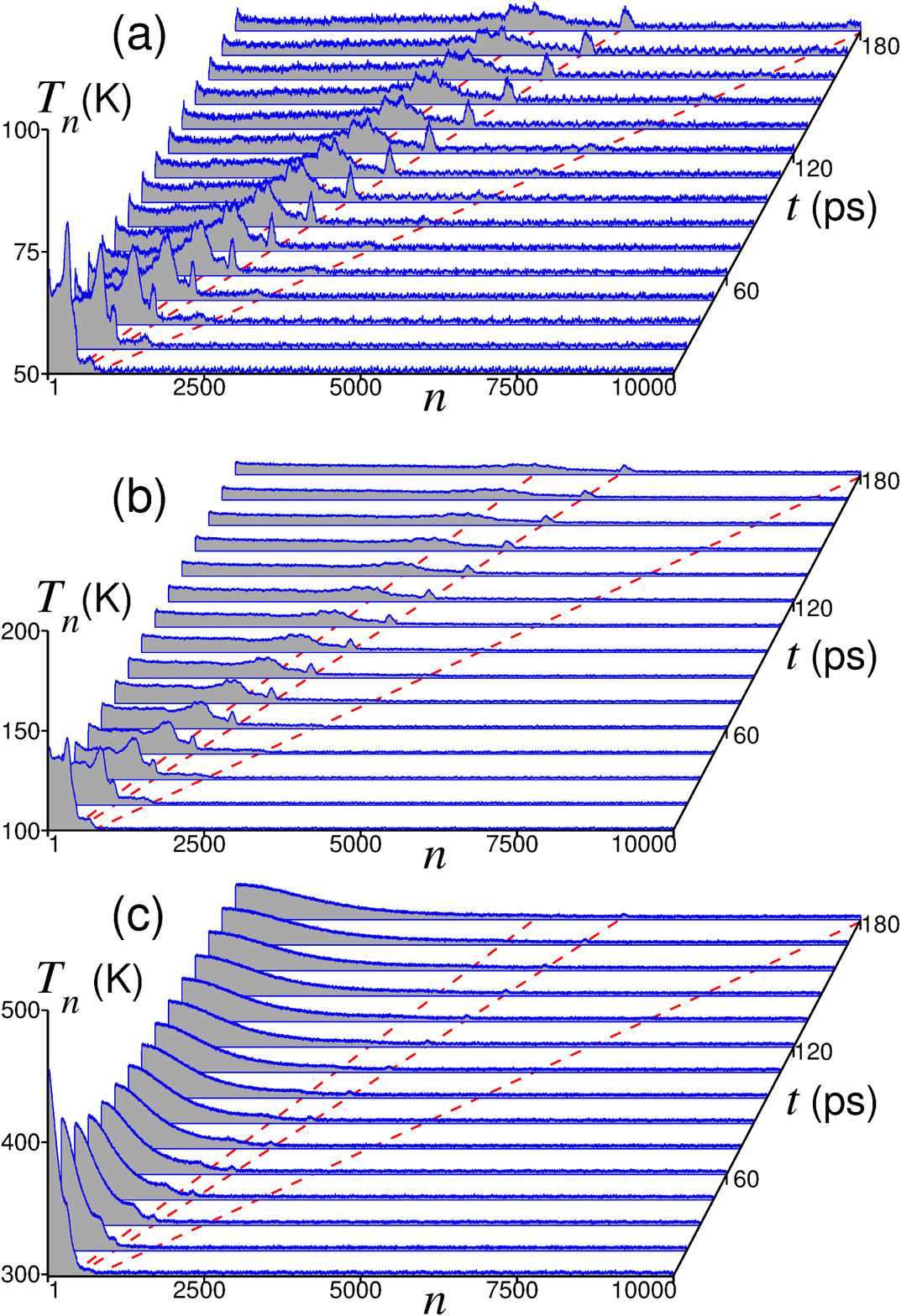}
\caption{
Dynamics of the temperature impulse along CNT with chirality index (6,6)
at its initial temperature $T_h=3T$, length $N_t=40$ and background temperature
(a) $T=50$, (b) 100 and (c) 300K.
The dependence of the temperature distribution along the CNT $T_n$ on the time $t$ is shown.
The dashed (red) lines shows trajectories for moving with constant speed $v=v_o$, $v_t$ and $v_l$.
}
\label{fig06}
\end{figure}
\begin{figure}[t]
\includegraphics[angle=0, width=1.0\linewidth]{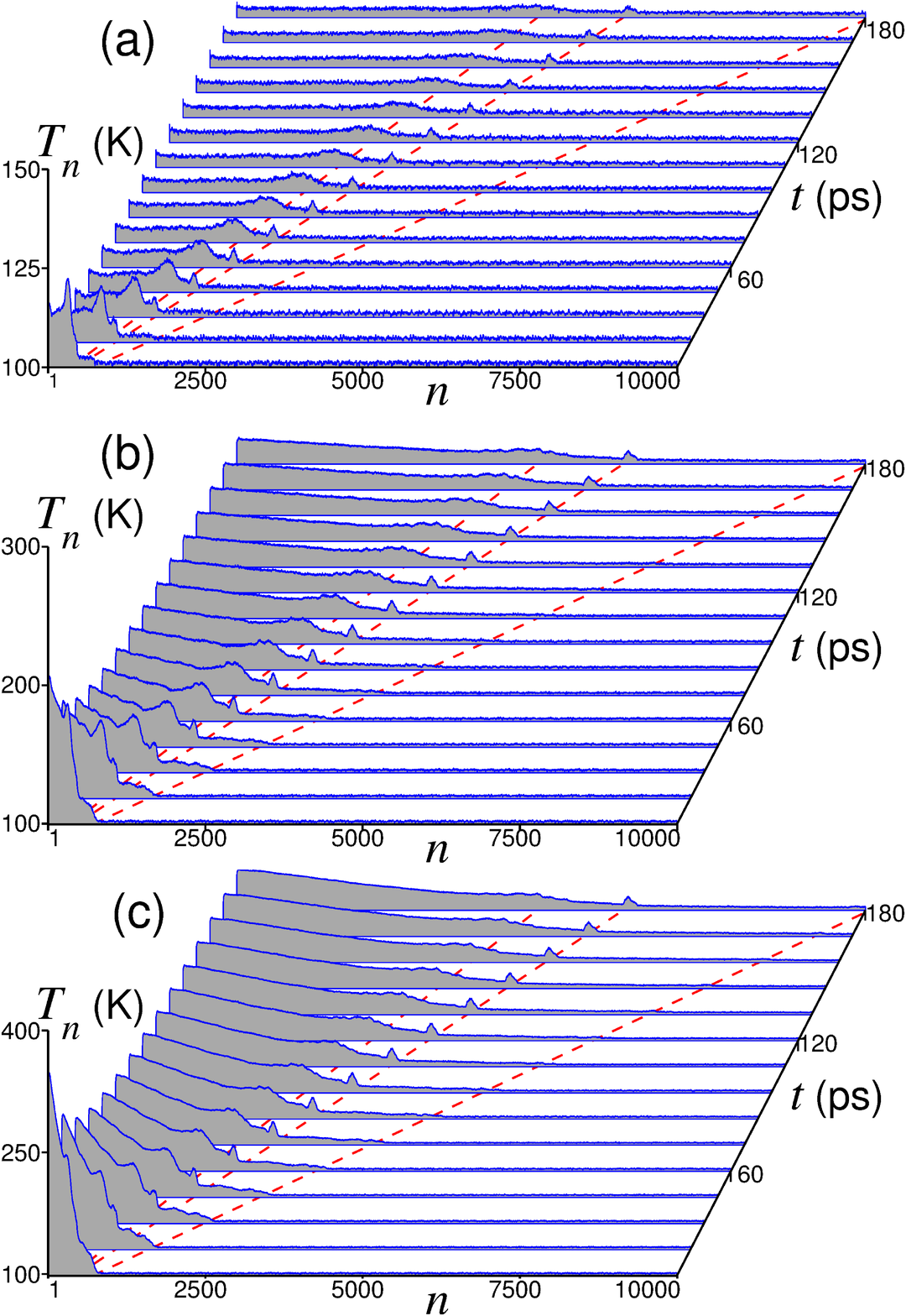}
\caption{
Dynamics of the temperature impulse along CNT (6,6) at its initial temperature
(a) $T_h=200$, (b) 500 and (c) 900 K ($T=100$K, $N_t=40$).
The dependence of the temperature distribution along the CNT $T_n$ on the time $t$ is shown.
The dashed (red) lines shows trajectories for moving with constant speed $v=v_o$, $v_t$ and $v_l$.
}
\label{fig07}
\end{figure}

\section{Dynamics of a thermal pulse}
\label{s5}
Let us simulate the propagations of a thermal impulse along carbon nanoribbon (CNR) and nanotube (CNT)
at different values of background temperature.
For this purpose, we take the finite CNR and CNT presented in Fig.~\ref{fig01} consisting of $N=10000$
transversal cells with fixed ends: $\dot{\bf x}_1\equiv {\bf 0}$, $\dot{\bf x}_N\equiv{\bf 0}$.

Then we take CNR (CNT) in the ground stationary state and thermalize it so that the first $N_t=40$
unit cells have a higher temperature than remaining cells.
For this purpose, we integrate Langevin's system of equations of motion with color noise
(\ref{f20}), (\ref{f21}) with the initial temperature distribution:
\begin{equation}
\{T_n=T_h\}_{n=1}^{N_t},~~~\{T_n=T\}_{n=N_t+1}^N,
\label{f30}
\end{equation}
where the temperature of the left edge is higher than the temperature of the main part of
CNR (CNT) ($T_h>T$).

After integrating the system of equations of motion (\ref{f20}), (\ref{f21}) during the time $t_1=5$~ns,
we will have the thermalized state of the system
\begin{equation}
\{{\bf x}_n(t_1),~\dot{\bf x}_n(t_1)\}_{n=1}^N,
\label{f31}
\end{equation}
in which all phonons of the nanoribbon (nanotube) will be thermalized according to their
quantum statistics (\ref{f12})
(with almost complete thermalization of low-frequency and partial thermalization of high-frequency phonons).
Next, we disable the interaction of the molecular system with the thermostat,
i.e. we already integrate the system of Hamilton equations (\ref{f24}) with the initial condition (\ref{f31}).
Using the formula (\ref{f28}), we will monitor the change of temperature profile along CNR (CNT) $\{T_n(t)\}_{n=1}^N$.
To increase the accuracy, the temperature profile is determined by $10^4$ independent
realisations of the initial thermalized state of the molecular system.
\begin{figure}[tb]
\includegraphics[angle=0, width=1\linewidth]{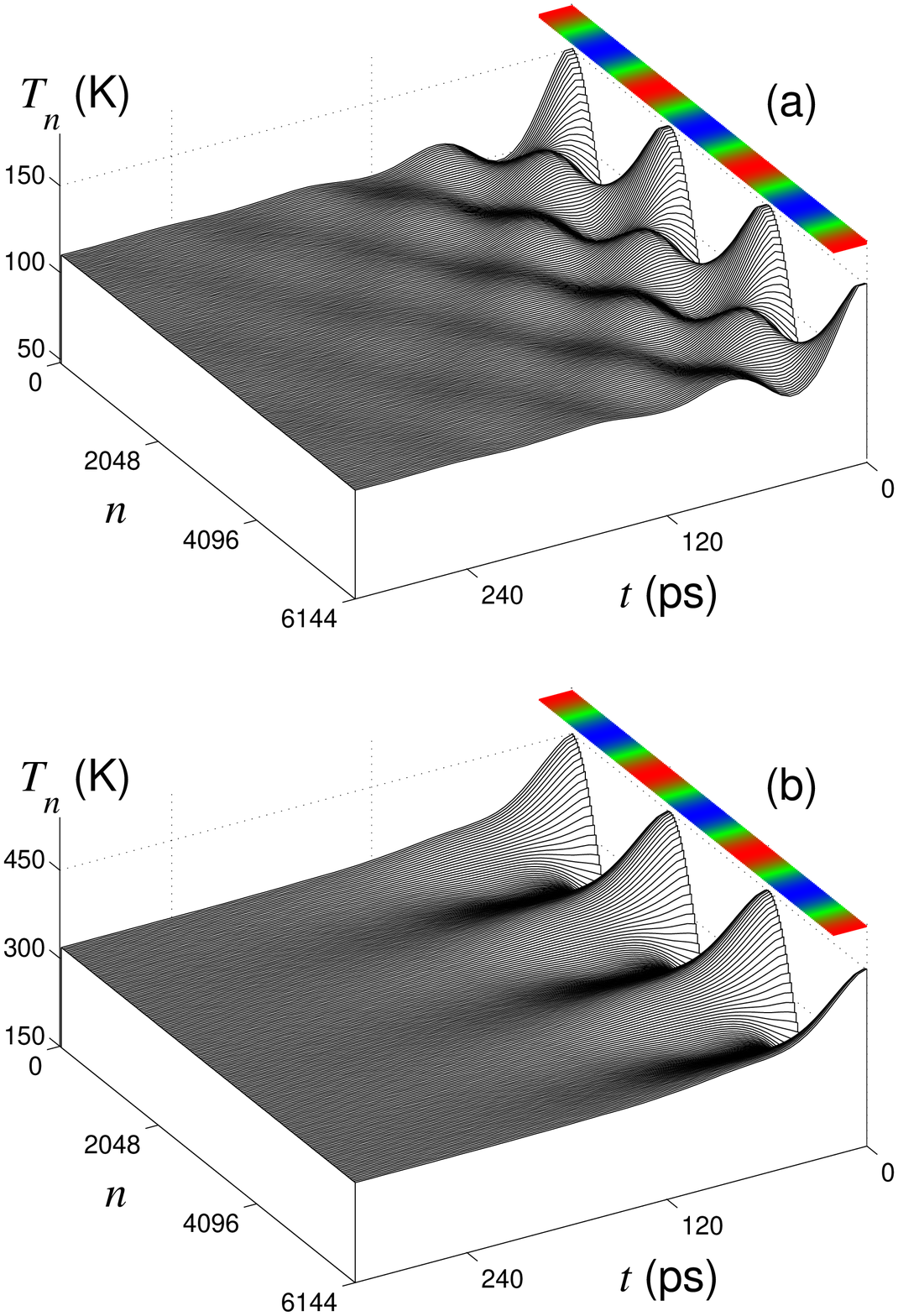}
\caption{
Relaxation of initial periodic thermal profile (temperature lattice) in the
nanoribbon with width $L_y=2.41$~nm for temperature (a)
$T=100$K ($\Delta T=50$K) and (b) $T=300$K ($\Delta T=150$K).
Dimensionless period of the profile $Z=2^{11}$, nanoribbon length $N=2^{13}$.
The temperature change on three periods of the profile is shown.
The initial temperature distribution in CNR is shown at the top
of each figure.
}
\label{fig08}
\end{figure}
\begin{figure}[tb]
\includegraphics[angle=0, width=1\linewidth]{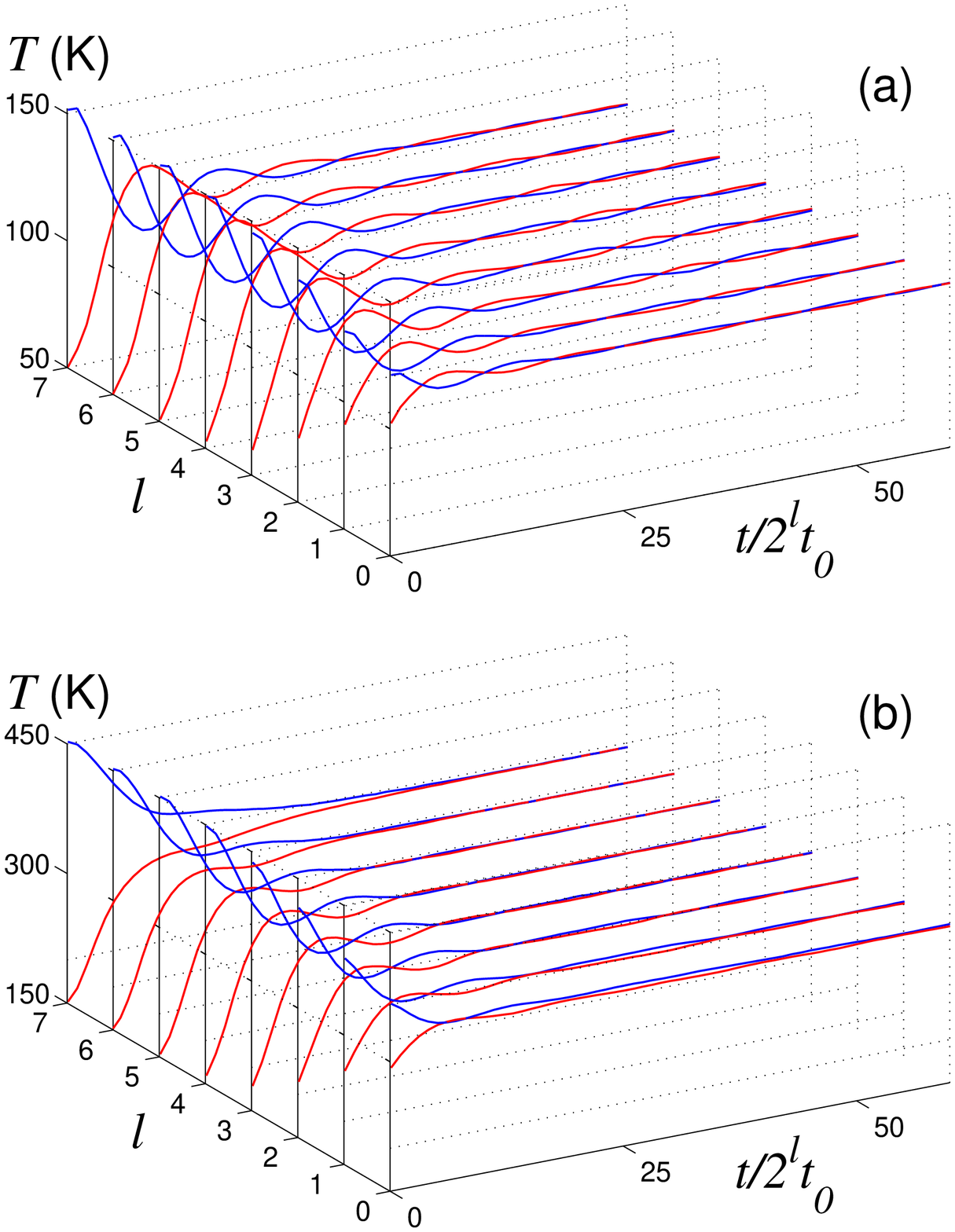}
\caption{
Evolution of the relaxation temperature profile in the zigzag nanoribbon with
width $L_y=2.41$~nm (length $N=2^{13}$)
for period $Z=2^{4+l}$ $(l=0,1,...,7)$ for (a) temperature $T=100$K ($\Delta T=50$K)
and (b) $T=300$K ($\Delta T=150$K).
Time dependencies of the mode maximum $T(1)$ [blue lines] and of
minimum $T(1+Z/2)$ [red lines] are depicted.
Time scales are differ for each curve in order to fit them into one figure, time $t_0=0.04$~ps.
}
\label{fig09}
\end{figure}

Modeling the propagation of an initial narrow temperature pulse ($N_t=40$, $T_h=3T$) along the nanoribbon
has shown that at low temperatures $50\le T\le 100$K, the thermal vibrations propagate
along the nanoribbon as a clearly visible wide wave the maximum of which moves at a speed
of $v=v_o$ -- see Fig.~\ref{fig05}~(a,b).
The value of the wave motion velocity suggests that the main role in its formation
is played by bending thermal phonons.
Therefore, here the velocity of the second sound corresponds to the maximum group velocity
of the optical bending phonons of the nanoribbon.
At a higher temperature $T>100$K, no temperature waves are formed, we see only a slow expansion
of the initial temperature impulse -- see Fig.~\ref{fig05}~(c).
Such dynamics is typical for the diffusion regime of heat transfer.

In the nanotube, the motion of heat waves is more pronounced -- see Fig.~\ref{fig06}~(a,b).
For $50\le T\le 150$K, two waves are clearly distinguished: a stable localized wave moving
at the speed of acoustic long-wave torsional phonons $v_t$ and a wider wave moving
at the maximum speed of optical bending phonons $v_o$.
The first wave corresponds to the ballistic regime of heat transfer, and the second wave
-- to the second sound (to hydrodynamic regime of heat transfer).
As the temperature increases, the heat waves become subtle.
At $T=300$K, the first (ballistic) wave becomes barely noticeable, and the second quickly disappears.
Here, the temperature propagates along the nanotube in the form of a diffusive expansion
of the initial impulse -- see Fig.~\ref{fig06}~(c).
Let us note that an increase in the amplitude of the initial temperature impulse $T_h$
does not lead to an increase in the proportion of energy carried by heat waves.
On the contrary, an increase in $T_h$ leads to the propagation of energy in the form
of a diffusive expansion of the initial temperature pulse even at a small value
of the background temperature $T$ -- see Fig.~\ref{fig07} and Fig.~\ref{fig06} (b).
\begin{figure}[tb]
\includegraphics[angle=0, width=1\linewidth]{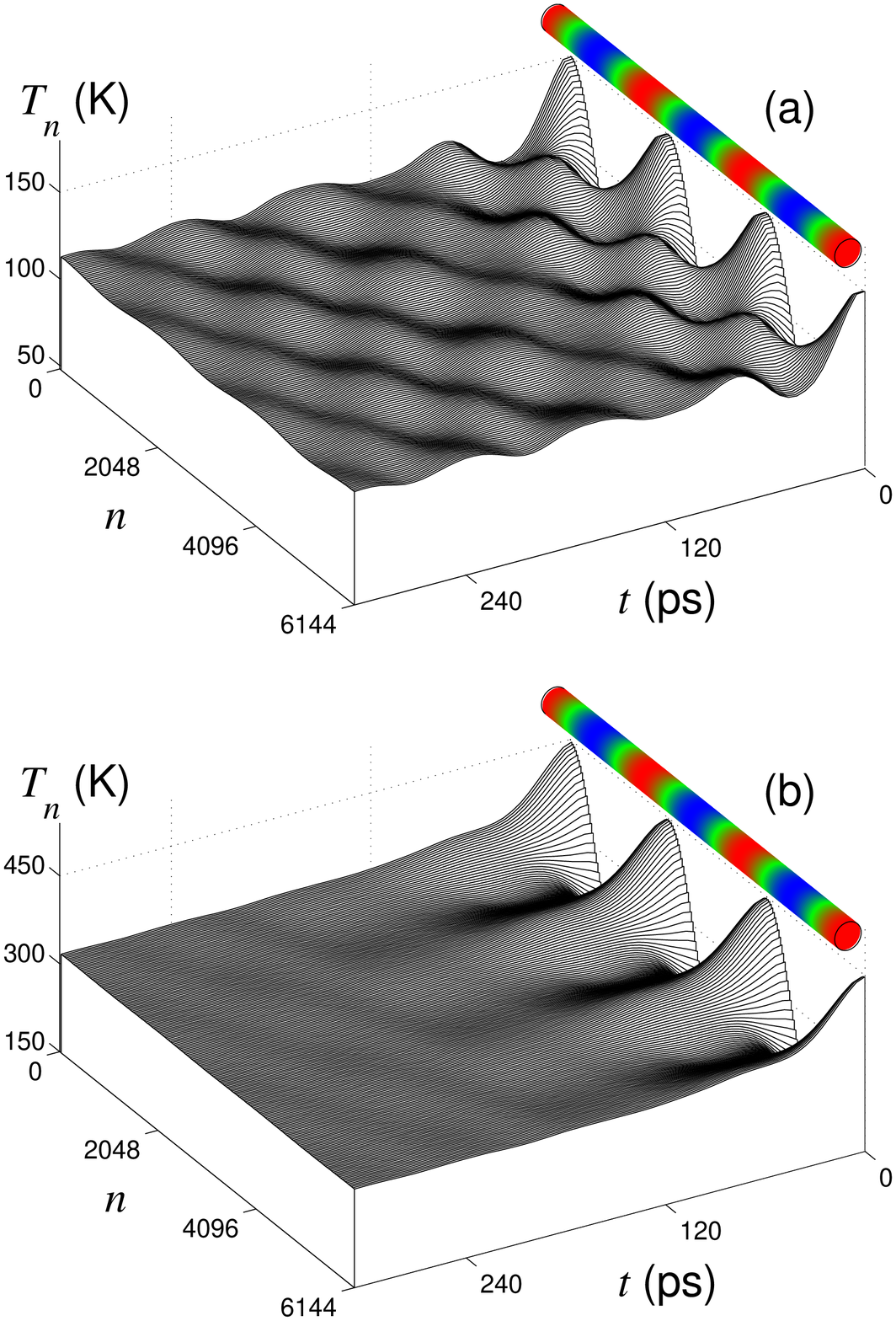}
\caption{
Relaxation of initial periodic thermal profile (temperature lattice) in the CNT
(6,6) for temperature (a) $T=100$K ($\Delta T=50$K) and (b) $T=300$K ($\Delta T=150$K).
The period of the profile $Z=2^{11}$, nanoribbon length $N=2^{13}$.
The temperature change on three periods of the profile is shown.
The initial temperature distribution in CNT is shown at the top of each figure.
}
\label{fig10}
\end{figure}
\begin{figure}[tb]
\includegraphics[angle=0, width=1\linewidth]{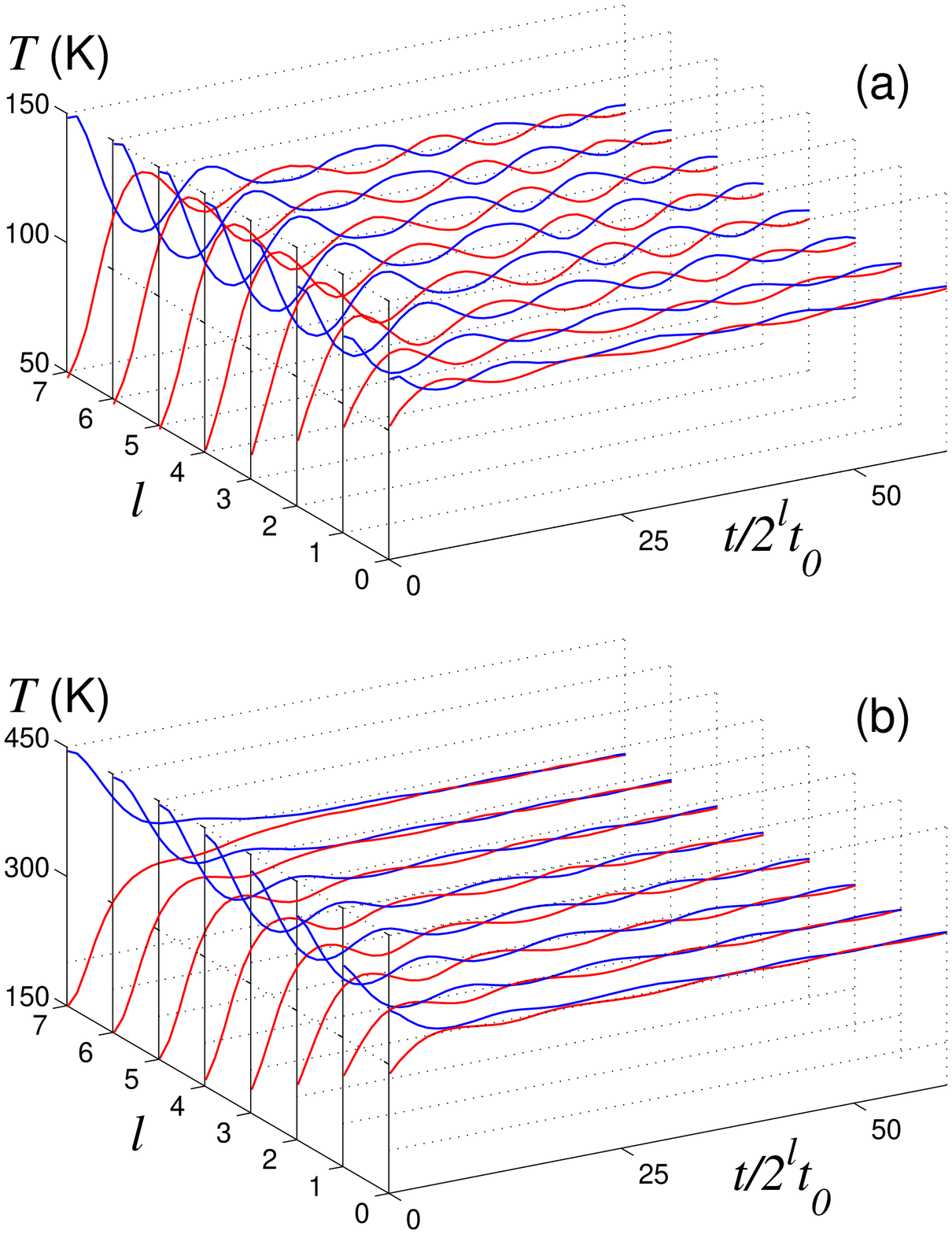}
\caption{
Evolution of the relaxation temperature profile in the CNT (6,6) (length $N=2^{13}$)
for period $Z=2^{4+l}$ $(l=0,1,...,7)$ for (a) temperature $T=100$K ($\Delta T=50$K)
and (b) $T=300$K ($\Delta T=150$K).
Time dependence of the mode maximum $T(1)$ [blue lines] and
minimum $T(1+Z/2)$ [red lines] are depicted.
Time scales are differ for each curve in order to fit them into one figure, time $t_0=0.04$~ps.
}
\label{fig11}
\end{figure}

\section{Relaxation of a periodic thermal lattice}
\label{s6}
The second sound can also be determined from the relaxation scenario of the initial
periodic sinusoidal temperature profile (from the relaxation of the temperature lattice).
Note that in the experimental works \cite{Huberman2019,Ding2022} the second sound
in graphite was determined from the analysis of relaxation of the temperature lattice.

Consider the finite nanoribbon (nanotube) of the length 2012~nm
(number of unit cells $N=2^{13}=8192$) with periodic boundary conditions.
In order to study the relaxation of the initial periodic temperature distribution
let us consider dynamic of CNR (CNT) with sinusoidal temperature profile
\begin{equation}
\{T_n=T+\Delta T\cos[2\pi(n-1)/Z]\}_{n=1}^N,
\label{f32}
\end{equation}
where $T$ is the average temperature, $\Delta T$ is the profile amplitude, and $Z=2^{4+l}$,
$l=0,...,7$ is dimensionless period of the profile.

To obtain the initial thermalized state, we numerically integrate the Langevin system of equations
with color noise (\ref{f20}), (\ref{f21}) with temperature distribution (\ref{f32})
during the time $t_1=5$~ps.
Then we use the resulting normalized state CNR (CNT) (\ref{f31}) as the starting point
for the Hamiltonian system equations of motion (\ref{f24}).
As a result of numerical integration of this system, using the formula (\ref{f28}), we have
the time dependence of the temperature distribution along the nanoribbon (nanotube) $\{T_n(t)\}_{n=1}^N$.
To increase accuracy, the results were averaged over $10^4$ independent realizations
of the initial thermalized state of the molecular system.

Let us take two values of the average temperature $T=100$, 300K and the lattice amplitude $\Delta T=0.5T$.
The results of numerical simulation of the relaxation of the temperature lattice in the nanoribbon
are shown in Fig.~\ref{fig08} and \ref{fig09}, and in the nanotube -- in Fig.~\ref{fig10} and \ref{fig11}.

As can be seen from Fig.~\ref{fig08}, at the profile period $Z=2^{11}$, damped periodic
fluctuations of the temperature profile occur in the nanoribbon at low temperature $T=100$K.
Such behavior corresponds to the hydrodynamic regime of heat transfer.
The speed of the temperature wave can be determined from the oscillation period $t_p$: $v(Z)=aZ/t_p$.
Damped profile fluctuations occur for all its considered periods $Z=2^4,2^5,...,2^{11}$
-- see Fig.~\ref{fig09}. At $T=300$K there are no periodic fluctuations. We only see a smooth spreading
of the initial temperature profile. This behavior is typical for the diffusion regime of heat transfer.

The typical behavior of the temperature profile in a nanotube is shown in Fig.~\ref{fig10}.
As can be seen from the figure, when the profile period is $Z=2^{11}$ and the temperature is $T=100$K,
slowly attenuating periodic profile oscillations occur in the nanotube.
This behavior corresponds to the hydrodynamic regime of heat transfer.
Profile fluctuations occur at all considered values of its period $Z=2^4,2^5,...,2^{11}$
-- see Fig.~\ref{fig11}.
At $T=300K$ low-amplitude profile fluctuations are noticeable only for $Z\le 2^9$.
Here, an increase in the profile period leads to a rapid transition of the profile dynamics
to the diffusion regime.
We can say that at $T=300$K, the second sound can manifest itself only at the lengths $L\le 2^9a=125.7$~nm.
\begin{figure}[tb]
\includegraphics[angle=0, width=1\linewidth]{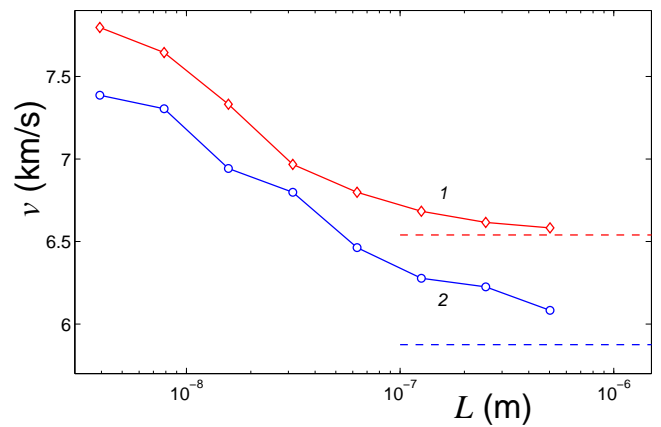}
\caption{
Dispersion of thermal periodic waves for $T=100$K.
Dependence of the wave velocity $v$ on its length $L=aZ$ for both CNT
and CNR (curves 1 and 2). Dashed lines mark the maximum values of the group velocities of the optical bending
phonons of CNT ($v_o=6.54$~km/s) and CNR ($v_o=5.87$~km/s), respectively.
}
\label{fig12}
\end{figure}

The dispersion of temperature waves at a temperature of $T=100 $K is shown in Fig.~\ref{fig12}.
With the increase of wavelength $L=aZ$, its velocity $v(Z)$ monotonically decreases and
in the limit tends to the maximum group velocity of optical bending phonons $v_o$.
Therefore, in carbon nanoribbons and nanotubes, the velocity of the second sound coincides with $v_o$,
which allows us to conclude that a high-temperature second sound occurs
primarily due to heat transfer by optical bending phonons.
\begin{figure}[tb]
\includegraphics[angle=0, width=1\linewidth]{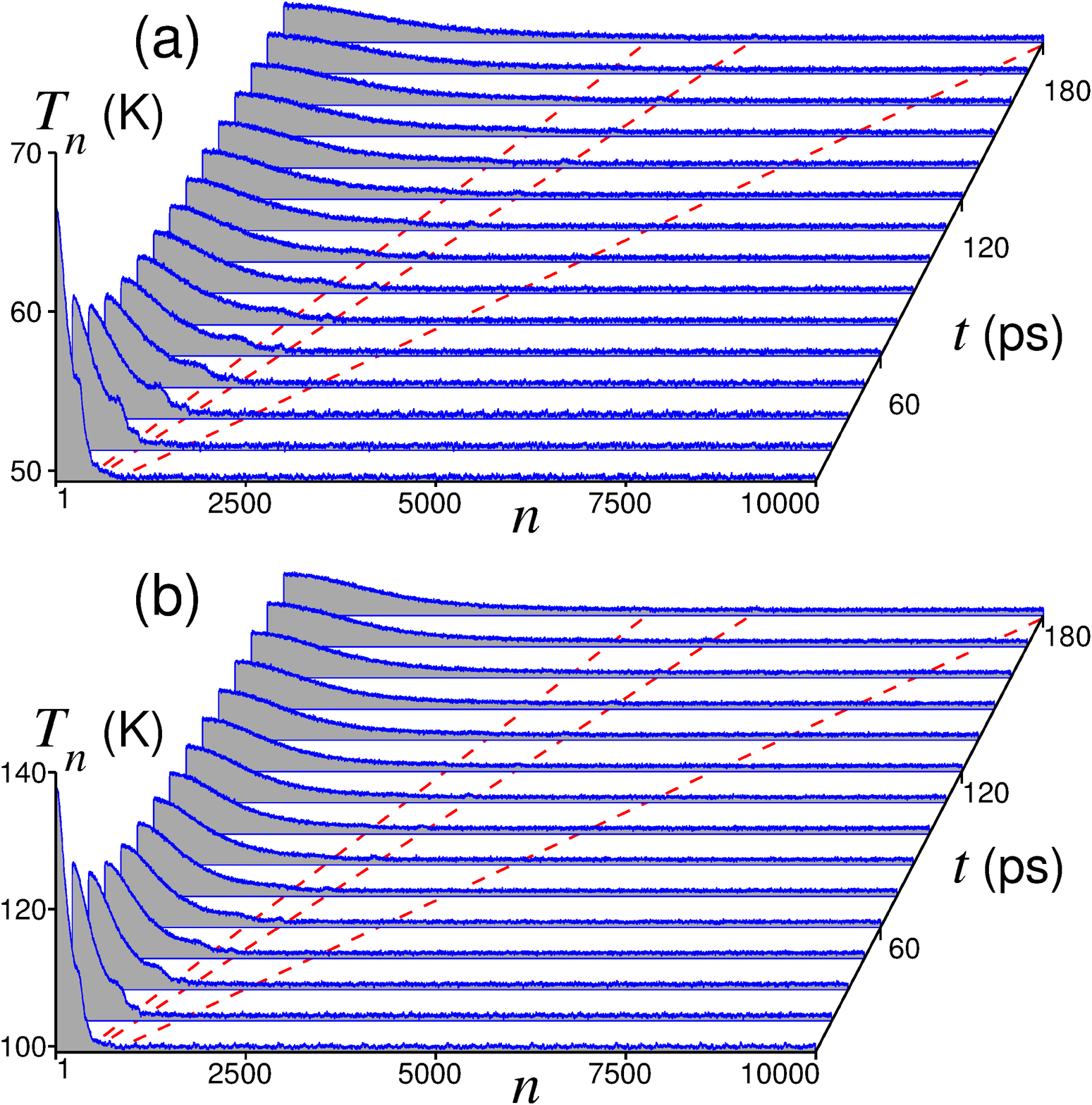}
\caption{
Dynamics of the temperature pulse along CNT (6,6) at its initial temperature $T_h=3T$,
length $N_t=40$ and background temperature (a)  $T=50$K and (b) $T=100$K.
The dependence of the temperature distribution along CNT $T_n$ on the time $t$ is shown.
The initial state of the temperature lattice is obtained using the classical
approximation (with full thermalization of all phonons).
The dashed (red) lines shows trajectories for moving at a constant speed $v=v_o$, $v_t$ and $v_l$.
}
\label{fig13}
\end{figure}
\begin{figure}[tb]
\includegraphics[angle=0, width=1\linewidth]{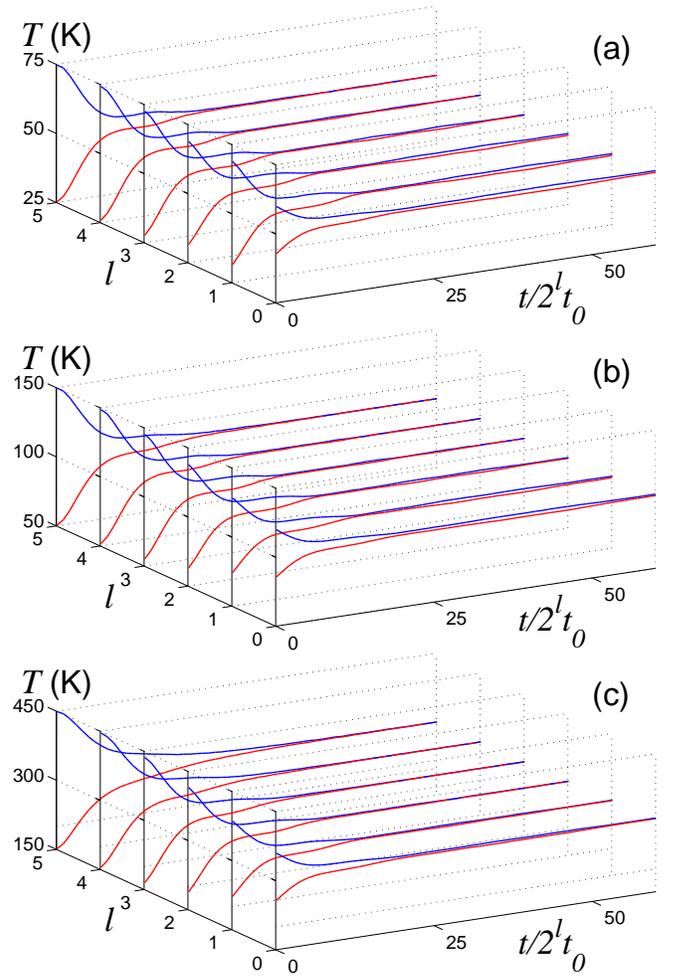}
\caption{
Evolution of the relaxation temperature profile in the CNT (6,6) (length $N=2^{13}$)
for period $Z=2^{4+l}$ $(l=0,1,...,5)$ for (a) temperature $T=50$K
(b) $T=100$K, and (c) $T=300$K (profile amplitude $\Delta T=T/2$).
Initial state of the temperature lattice is obtained using the classical
approximation (with full thermalization of all phonons).
Time dependence of the mode maximum $T(1)$ (blue lines) and
minimum $T(1+Z/2)$ (red lines) are depicted.
Time scales differ for each curve in order to fit them into one figure, time $t_0=0.04$~ps.}
\label{fig14}
\end{figure}

\section{Comparison with classical molecular dynamics}
\label{s7}

Our simulation of the dynamics of temperature impulses and relaxation of periodic temperature
lattices has shown that by using semi-quantum molecular dynamic, i.e., by
taking into account the quantum statistics of thermal phonons, at a temperature of $T=100$K and below,
a hydrodynamic heat transfer regime, the indicator of which is the second sound, is observed in
carbon nanotubes and nanoribbons.
The increase in temperature leads to a transition from the hydrodynamic to the diffusion regime
of heat transfer. At room temperature $T=300$K, there is no second sound.
This conclusion is in agreement with the results of the experimental work \cite{Huberman2019},
in which fast, transient measurements of the thermal lattices showed the existence of a second
sound in graphite at temperatures of $85<T<125$K.
To assess the importance of taking into account the quantum statistics of thermal phonons,
we will check the existence of a second sound using the classical method of molecular dynamics,
in which all phonons, regardless of temperature and their frequency, are equally thermalized.

In the classical method of molecular dynamics, to obtain a thermalized initial state of CNT (CLR) (\ref{f31}),
it is necessary to integrate a system of Langevin equations with white noise (\ref{f13}), (\ref{f14})
with a temperature distribution (\ref{f30}) to simulate the motion of the temperature impulse,
and with a distribution (\ref{f32}) to simulate the relaxation of the temperature lattice.

The simulation of the motion of the temperature impulse has shown that with the same thermalization
of all phonons, the wave-like motion of temperature in CNT and CNR does not occur even
at low temperatures -- see Fig.~\ref{fig13}.
For example, at $T=50$, 100K, the transfer of the temperature impulse along the nanotube occurs
in the form of its slow diffusion expansion, which is sharply different from the scenario obtained
using the semi-quantum method -- see Fig.~\ref{fig06}.
Modeling of the relaxation of the temperature lattice also shows that when using the classical
method of molecular dynamics (with full thermalization of all phonons),
temperature waves are not formed even at low temperatures, there is only a slow diffusion spreading
of the temperature lattice -- see Fig.~\ref{fig14}.
Thus, the classical molecular dynamics does not allow to simulate temperature waves
experimentally observed in graphite \cite{Huberman2019,Ding2022}.
Therefore, it is fundamentally important to take into account the quantum statistics of thermal phonons
for modeling the second sound in quasi-one-dimensional and two-dimensional molecular systems.

Note that graphene has a very high Debye temperature $T_D=\hbar\omega_m/k_B=2300$K,
where $\omega_m=1600$~cm$^{-1}$ is the maximum frequency of the phonon spectrum.
Therefore, without taking into account the quantum statistics of phonons,
it is impossible to obtain reliable results for graphene nanoribbons and nanotubes.

\section{Concluding remarks}
\label{s8}

Our numerical studies of heat transport in low-dimensional carbon structures has revealed that, 
in order to explain the observation of high-temperature second sound in crystalline graphite, 
it is critically important to taking into account quantum statistics of thermal phonons.  
In contrast, the classical method of molecular dynamics with full thermalization of all phonons 
does not allow simulating correctly second sound at temperatures $85\div 125$K observed in experiment~\cite{Huberman2019}.
When the quantum statistics is taken into account, only nonzero phonons with the average energy $k_BTp(\omega,T)$,
where the density distribution of energy phonons $p$ depends on the temperature according to the 
formula (\ref{f12}), can participate in heat transport.  By decreasing the temperature, only 
low-frequency phonons with frequencies $\omega<k_BT/\hbar$ remain fully thermalized.

From the form of the dispersion curves for nanoribbons and nanotubes (see Figs. \ref{fig03} and \ref{fig04}),
it follows that at $T=300$K, almost all phonons with frequencies $\omega<300$~cm$^{-1}$ remain
thermalized and participate in heat transfer. 
For low temperatures $T<50$K, only low-frequency long-wave phonons participate in heat transfer. 
For a nanotube, these are two acoustic branches (longitudinal and torsional acoustic phonons having
velocity $v_l$ and $v_t$) and two optical branches leaving the zero point (with wave number 
$q\searrow 0$, their phase velocity tends to zero). 
Therefore, for very low temperatures, acoustic long-wave phonons make the main contribution to 
heat transfer, thus defining the ballistic regime of heat transfer. 
For high temperatures, the participation of all phonons in heat transfer leads to a diffusion regime.
For intermediate temperatures $50\div 150$K, only low-frequency phonons participate in heat transfer,
a large part of them will be accounted for by bending phonons -- see Fig.~\ref{fig03}(b).

For a nanotube, optical bending waves have a maximum group velocity $v_o=6.54$~km/s --
the velocity of bending phonons with dimensionless wave numbers $q\in [0.05,0.5]$ (see Fig.~\ref{fig04}).
For a nanoribbon, this velocity is $v_o=5.87$~km/s. For temperatures $50\div 150$K, bending phonons
with these wave numbers make the main contribution to heat transfer. 
In the simulations, the motion of these phonons appears as the motion of a temperature maximum 
at velocity $v=v_0$ -- see Figs.~\ref{fig05} and \ref{fig06}.
Therefore, we can conclude that second sound observed in graphite at liquid nitrogen temperatures
corresponds primarily to the motion of optical bending vibrations.

Second sound will be absent for very low temperatures $T<20$K, since in this case
bending phonons with velocities close to $v_o$ will no longer participate in heat transfer.
For high temperatures $T>200$K, second sound becomes faintly noticeable, since many phonons here 
already participate in heat transfer, 
and the contribution from the motion of low-frequency bending phonons will remain relatively small.

Thus, we come to the conclusion that, for carbon nanoribbons and nanotubes, high-temperature second
sound is caused by low-frequency bending phonons. The existence of such phonons follows from the one-dimensionality
or two-dimensionality of the molecular systems. Therefore, we should expect second sound in such 
two-dimensional systems as graphene (graphite) \cite{Huberman2019,Ding2022}, 
hexagonal boron nitride (h-BN), and in quasi-one-dimensional systems such as graphene and h-BN nanotubes,
linear cumulene macromolecule \cite{Melis2021}, and planar zigzag polyethylene. 
In quasi-one-dimensional systems, second sound manifests itself more strongly than in quasi-two-dimensional
systems, since bending vibrations of the former make a greater contribution to heat transfer.
The speed of second sound in such molecular structures can be estimated as the maximum
group velocity of bending optical phonons whose dispersion curve split off the zero point.
Our results suggest that the velocity of second sound in graphene (in graphite) is about 6 km/s.

Finally, we notice that, because bending vibrations have nonzero dispersion \cite{Tornatzky2019,Mahrouche2022},
there are no optical bending phonons in two-dimensional layers of MoS$_2$, MoSe$_2$, WS$_2$, and WSe$_2$
where the molecular layers have a finite thickness (such as flat corrugated structures).
Therefore, high-temperature second sound should not be observed in all such structures. 
Instead, only a ballistic regime should be expected, with a transition to a diffusive heat 
transfer regime with a growth of temperature.

\begin{center}
{\bf ACKNOWLEDGMENTS}
\end{center}

A.S. acknowledges the use of computational facilities at the Interdepartmental Supercomputer 
Center of the Russian Academy of Sciences. 
Y.K. acknowledges a support from the Australian Research Council (grants DP200101168 and
DP210101292).

\end{document}